\documentclass[twocolumn,tighten]{aastex6}

\usepackage{graphicx}	
\usepackage{amsmath}	
\usepackage{amssymb}	
\usepackage[hyphenbreaks]{breakurl}
\usepackage{upgreek}

\newcommand\chandra{{\it Chandra}}
\newcommand\xmm{{\it XMM-Newton}}
\newcommand\suzaku{{\it Suzaku}}
\newcommand\nustar{{\it NuSTAR}}
\newcommand\swift{{\it Swift}}

\newcommand\ergsec{{\rm~erg\ s}^{-1}}

\begin{document}

\title{Exploring the Spectral Variability of the Ultra-luminous X-ray source M81 X--6 with {\it Suzaku} and {\it XMM-Newton}}
\shorttitle{Variability of ULX M81 X--6}

\author{V. Jithesh and Ranjeev Misra}
\affil{Inter-University Centre for Astronomy and Astrophysics, Post Bag No. 4, Ganeshkhind, Pune 411007, India; vjithesh@iucaa.in}
\shortauthors{Jithesh and Misra}

\begin{abstract}

We present X-ray spectral variability studies of the ultra-luminous X-ray source (ULX) M81 X--6 using {\it Suzaku} and {\it XMM-Newton} observations performed during 2001--2015. The spectra were first fitted by a standard multi-temperature disk and a thermal Comptonization component which revealed spectral variability where the primary distinction is the change in the optical depth of the Comptonizing component, similar to what has been observed for other ULXs. We also fitted the spectra with a general relativistic accretion disk emission and a power-law component and found that it can reproduce a large part but not all of the spectral variability of the source. The parameters for the black hole mass and spin were found to be degenerate, but the high spin and larger mass ($20-100\,\rm M_\odot$) solutions provided near Eddington accretion rates consistent with the assumptions of the model. The spectral variation is found to be driven by accretion rate changes leading to three different spectral classes. Thus, our results suggest the possibility of a dominant relativistic disk emission component for some of the spectral states of the source.

\end{abstract}


\section{Introduction} 
\label{sec:intro}

Ultra-luminous X-ray sources (ULXs) are a class of extragalactic, compact, non-nuclear X-ray sources with X-ray luminosities in the range of $10^{39-41} \ergsec$ \citep[see][for the reviews on ULXs]{Fen11, Kaa17}. Such luminosities exceed the Eddington limit for accretion of pure hydrogen on to a typical $10\,\rm M_{\odot}$ black hole (BH). Early studies with phenomenological two-component model suggest a cool accretion disk for ULXs \citep{Mil03}, which was interpreted as the evidence for the presence of intermediate-mass black holes (IMBHs) of mass $\sim 10^{2} - 10^{4}\,\rm M_{\odot}$ \citep{Col99, Mil04a}. Unfortunately, the detailed studies of ULXs failed to identify any signatures of standard thin accretion disk spectrum that may be associated with IMBHs \citep{Rem06, Rob07}. However, in the most luminous ULXs ($L_{\rm X} \gtrsim 5 \times 10^{40} \ergsec$), IMBHs may be present \citep{Far09, Sut12}. A cool disk and a high-energy curvature have been observed in several high-quality ULXs spectra \citep{Rob05, Sto06, Gon06} and these peculiar spectral features suggest a new super-Eddington state, referred as {\it ultraluminous state} \citep{Gla09}, for ULXs. Thus, majority of the ULXs are likely powered by accretion on to stellar-mass BHs (StMBHs; $M_{\rm BH} < 20\,\rm M_{\odot}$) with super-Eddington rates. Alternatively, some ULXs can contain slightly larger mass BHs, with a range of $20-100\,\rm M_{\odot}$ \citep[massive BHs;][]{Zam09, Fen11}. In such cases, the ULXs can accrete at Eddington or near Eddington rates to produce the observed high X-ray luminosities. 

One of the major breakthroughs in this field is the detection of pulsation from the ULX M82 X--2 \citep{Bac14}. This result confirms the presence of a neutron star (NS) in these high luminosity objects. After this discovery, few other sources (NGC 7793 P13, NGC 5907 ULX1, and NGC 300 ULX) have shown pulsation and hence demonstrate that the high luminosity of ULXs can even come from an NS system \citep{Isr17a, Isr17b, Car18}. Therefore, ULXs may represent an heterogeneous class of objects, which include the IMBHs, massive BHs (MBHs), StMBHs and NS systems.

ULXs are variable sources and exhibit different spectral shapes, which are related to different spectral states possibly achieved by state transitions \citep{Kub01, God09, Dew10}. ULXs IC 342 X--1 and X--2 showed two distinct spectral shapes in 1993 and 2000 {\it Advanced Satellite for Cosmology and Astrophysics} ({\it ASCA}) observations, which are interpreted as the canonical transition between low/hard and high/soft states \citep{Kub01}. The analysis of two {\it XMM-Newton} observations of the ULX NGC 1313 X--1 revealed that the source was in two different spectral states and the spectral shapes are varied between the two observations \citep{Dew10}. Such states can be explained by invoking different accretion disk geometry, where the low flux state is consistent with truncated disk model and the hot corona covers the cold accretion disk in the high flux state. A detailed study with all \xmm{} observations also showed different spectral states (``very-thick'' and ``thick'' state) for NGC 1313 X--1 and X--2, and these states exhibit different spectral shapes \citep{Pin12}. In the very high optical depths (``very-thick'' state), the spectra are bell-shaped with a clear curvature at high energies ($\sim$ 3--4 keV) , while in the ``thick'' state, the spectra are steeper with no strong evidence of curvature \citep[see Figure 2 in][]{Pin12}. Thus, these sources have two well-defined states in their respective observations and thereby distinct spectral shapes.

ULXs spectra were described by different spectral models that include the simple models as well as complex phenomenological models. Spectral modeling with simple models such as multi-color disk blackbody (MCD) plus power-law are qualitatively similar to the model used for Galactic X-ray binaries. Such model description was used in the past to reproduce the low counting statistics spectra of ULXs and returned a low inner disk temperature in the range of $\sim 0.1 - 0.4$ keV with the power-law index of $\sim 1.5-2.5$ \citep{Mil03, Fen05}. The high luminosity and cool accretion disk component were erroneously considered as supportive evidence for the accretion onto IMBHs \citep{Mil03, Mil04}. On the other hand, the high counting statistics X-ray spectra of bright ULXs can be well fitted with different physical models, which suggest the presence of an optically thick corona, a fast ionized outflow and/or a slim disk \citep{Wat01, Sto06, Gon06}. The analysis of highest quality {\it XMM-Newton} spectra of several ULXs confirmed the existence of peculiar spectral features such as soft excess and broad curvature at high energy, which are almost ubiquitous in the high-quality ULX Spectra \citep{Gla09}. Such high-quality spectra can be described in terms of disk plus Comptonized corona models, which suggests a cool, optically thick corona for the majority of ULXs \citep{Gla09, Vie10, Pin12, Pin14}.

M81 X--6 (NGC 3031 X--11) is a ULX in the nearby spiral galaxy M81 at a distance of 3.63\,Mpc~\citep{Fre94}. The source was discovered by the {\it Einstein} observatory \citep{Fab88} and observed to be in the steady flux state throughout its observations. This source was also detected by the {\it ROSAT} and reported a change in the intensity by a factor of 2 over 6 day period in 1993 November \citep{Imm01}. {\it ASCA} spectra of the source taken during the period 1994--1997 were successfully described by the MCD emission arising from optically thick standard accretion disks around black holes \citep{Mak00, Miz01}. The high luminosity derived from the model suggests a high BH mass for the source, while the inferred temperature, in the range of 1--2 keV, is incompatible with the high mass. A rapidly spinning BH for M81 X--6 would be an alternative solution for this issue \citep{Miz01}. It was suggested based on X-ray spectral modeling of M81 X--6 spectra from the {\it Chandra} and {\it ASCA} observations that the system contains a $18\,\rm M_{\odot}$ BH \citep{Swa03}. In addition, an optical counterpart has been identified for M81 X--6 using the {\it Hubble Space Telescope} ($HST$) observations and the optical properties are consistent with an early-type main-sequence star of spectral class O8 V with a mass of $23\,\rm M_{\odot}$ \citep{Liu02}. The X-ray spectral study based on \suzaku{} and \xmm{} observations of M81 X--6 showed a rollover at $\sim 3$ keV in the broadband spectra (0.6--30 keV) with no additional high-energy PL component that would significantly contribute to the X-ray emission \citep{Dew13}. These results clearly demonstrate that the source exhibits unusual spectral states, which are significantly different from the state observed in BH X-ray binaries and active galactic nuclei (AGNs). Using {\it Swift} long-term monitoring observations of M81 X--6, \citet{Lin15} searched for periodic signals from the source and identified two peaks near 110\,d and 370\,d in the Lomb-Scargle periodogram (LSP). The observed peaks are not statistically very significant and are inconsistent with the prediction of \citet{Swa03}, where the estimated orbital period of the system is 1.8\,d. Thus, such large periods may probably be associated with the super-orbital period of the system \citep{Lin15}. 

\citet{Hui08} presented a sample study of six ULXs, including M81 X--6, using relativistic disk models, {\tt kerrbb} and {\tt bhspec}. Using single \xmm{} observation conducted in 2004 (with MOS spectra only), authors suggest a MBH in the range of $33-85\,\rm M_{\odot}$ for M81 X--6. Both models indicate that the source can accrete at values close to the Eddington limit and the spin value obtained from this study is consistent with maximally spinning BH with an inclination angle of $50-70^{\circ}$. In the present work, we study the long-term spectral variability of M81 X--6 using all archival \xmm{} and \suzaku{} observations conducted during 2001--2015. The observations and data reduction method is described in the \S 2. The analysis and results are presented in \S 3. The discussion is presented in \S 4, and in \S 5 we summarize our results. 

\section{Observations and Data Reduction}

We used the on-axis \suzaku{} and \xmm{} observations of M81 and Ho IX X--1, which cover target M81 X--6 in the field-of-view, for this study. The details of the observations are given in Table~\ref{sample}.

\begin{table}
\tablecolumns{5}
\setlength{\tabcolsep}{10.0pt}
\tablewidth{320pt}
	\caption{Observation log in the chronological order}
 	\begin{tabular}{@{}cccc@{}}
	\hline
	\hline
\colhead{Data} & \colhead{ObsID} & \colhead{Date} & \colhead{Exposure} \\
     	       &		 & 		  & (ksec) \\
\hline

X1 & 0111800101 & 2001 Apr 22 & 132.7 \\
X2 & 0111800301 & 2001 Apr 22 & 8.0 \\
X3 & 0112521001 & 2002 Apr 10 & 10.7 \\
X4 & 0112521101 & 2002 Apr 16 & 11.5 \\
X5 & 0200980101 & 2004 Sep 26 & 119.1 \\
S1 & 701022010 & 2006 May 08 & 103.5 \\
X6 & 0657802001 & 2011 Mar 24 & 27.5 \\
S2 & 906004010 & 2011 Sep 15 & 45.7 \\
X7 & 0657801601 & 2011 Apr 17 & 21.1 \\
X8 & 0657801801 & 2011 Sep 26 & 25.4 \\ 
X9 & 0657802201 & 2011 Nov 23 & 23.9 \\
X10 & 0693850801 & 2012 Oct 23 & 14.1 \\
X11 & 0693850901 & 2012 Oct 25 & 14.0 \\
X12 & 0693851001 & 2012 Oct 27 & 13.9 \\
X13 & 0693851701 & 2012 Nov 12 & 9.9 \\
X14 & 0693851801 & 2012 Nov 14 & 13.8 \\
X15 & 0693851101 & 2012 Nov 16 & 13.3 \\
S3 & 710017010 & 2015 May 18 & 97.7 \\

\hline
\end{tabular} 
\tablecomments {X and S indicate {\it XMM-Newton} and {\it Suzaku}, respectively.}
\label{sample}
\end{table}


\subsection{{\it XMM-Newton} Observations}

The \xmm{} observations conducted between 2001 and 2012 were obtained from the \xmm{} science archive\footnote{\url{http://nxsa.esac.esa.int/nxsa-web/\#home}}. The archival data were reduced using the standard tools available in the \xmm{} Science Analysis Software ({\sc sas}; version 14.0). We used the \xmm{} European Photon Imaging Camera (EPIC) pn \citep{Str01} and Metal Oxide Semiconductor \citep[MOS;][]{Tur01} instruments data for the analysis, and reduced the data using {\sc epchain} and {\sc emchain} tools, respectively. We extracted the full-field background light curve in the 10--12 keV energy band from both instruments and removed the high particle flaring background by applying the appropriate count rate cut-off criteria, thereby creating the good time intervals file for each observation. Three observations, X2, X6 and X7 in Table~\ref{sample}, considered in this analysis are affected by high amplitude particle flaring background. The available exposure time after filtering is too low to get a good quality spectrum in these observations, and thus we decided to exclude these observations from the analysis. The pn event files were filtered using the flag expression $\#XMMEA\_EP$ and PATTERN $ \le 4$ (single and double events), while for the MOS data, we used the $\#XMMEA\_EM$ flag and predefined patterns of 0--12. 

Circular regions of radius range 25--40 arcsec centered at the source position \citep[RA= 09:55:32.9, Dec.= +69:00:33.3, equinox J2000.0;][]{Gla09} were used to extract the source events. The different extraction radii were used to avoid the CCD chip-gaps. The background region was selected from a source-free region close to the source (from the same CCD where the source region was selected) with radius same as that of the source. We used the {\sc sas} task {\sc especget} to extract the source and background spectra along with the ancillary response file (ARF) and redistribution matrix file (RMF). We note that in some observations the target resides on the chip-gaps or close to the edge of the gap in the pn/MOS instruments. The X-ray events in the chip-gap are less accurately energy-calibrated and the adjacent events have uncertain patterns due to the possible charge loss in the gap. Such events have poor spectral calibration and can be excluded from the data with the `FLAG = 0' selection criterion \citep[see][for more details]{Dew10}. This exclusion reduces the flux significantly in these spectra, however, the {\sc sas} task {\sc arfgen} corrects the flux loss\footnote{\url{https://xmm-tools.cosmos.esa.int/external/xmm\_user\_support/documentation/sas\_usg/USG/}}. The \xmm{} observations conducted in October 2012 (X10--X12) and November 2012 (X13--X15) are separated by two days and gave low statistic spectra from individual observations. We checked the hardness ratios of the source in the individual observations using the count rates from soft (0.3--1 keV), medium (1--2 keV) 
and hard (2--6 keV) energy bands, and found that there is no major spectral variations occurred during these observations. Thus, we decided to add these spectra to get moderate quality spectra (A1 and A2; see Table \ref{chidof}). All the extracted spectra were then binned with a minimum count of 20--100 per bin, using {\sc grppha}, depending upon the quality of the data.

\begin{table*}
\tablecolumns{8}
\setlength{\tabcolsep}{12.0pt}
\tablewidth{320pt}
	\caption{The $\chi^2/\rm d.o.f$ for different models}
 	\begin{tabular}{@{}ccccccccc@{}}
	\hline
	\hline
\colhead{ObS ID} & \colhead{Model 1} & \colhead{Model 2} & \colhead{Model 3} & \colhead{Model 4} & \colhead{Model 4} & \colhead{Model 4} & \colhead{Model 4} \\
& & & & & \colhead{$a^{*}=0$} & \colhead{$a^{*}=0.9$} & \colhead{$a^{*}=0.999$} & \\

\hline

X1 & $719.9/405$ & $499.7/403$ & $465.0/402$ & $508.6/402$ & $511.0/403$ & $522.7/403$ & $523.4/403$ \\ 
X3 & $206.2/196$ & $184.3/194$ & $185.4/193$ & $184.5/193$ & $184.5/194$ & $184.8/194$ & $184.9/194$ \\ 
X4 & $240.6/217$ & $223.0/215$ & $222.9/214$ & $223.1/214$ & $223.2/215$ & $223.5/215$ & $223.2/215$ \\ 
X5 & $353.6/304$ & $277.2/302$ & $277.0/301$ & $276.0/301$ & $276.9/302$ & $276.6/302$ & $276.0/302$ \\ 
S1 & $509.2/273$ & $317.6/271$ & $313.5/271$ & $312.4/270$ & $315.4/271$ & $312.8/271$ & $316.7/271$ \\   
S2 & $224.8/232$ & $211.2/230$ & $212.2/230$ & $209.7/229$ & $211.2/230$ & $209.8/230$ & $210.1/230$ \\   
X8 & $196.7/195$ & $175.5/193$ & $169.0/192$ & $176.7/192$ & $176.9/193$ & $178.4/193$ & $178.8/193$ \\ 
X9 & $325.9/196$ & $168.5/194$ & $168.8/193$ & $170.9/193$ & $168.3/194$ & $168.4/194$ & $168.5/194$ \\
A1 & $216.5/227$ & $181.0/225$ & $178.8/224$ & $180.9/224$ & $181.0/225$ & $180.9/225$ & $181.4/225$ \\
A2 & $237.7/224$ & $223.3/222$ & $224.2/221$ & $222.9/221$ & $223.2/222$ & $223.1/222$ & $222.9/222$ \\  
S3 & $593.4/208$ & $189.4/206$ & $191.1/205$ & $189.4/205$ & $189.4/206$ & $189.5/206$ & $189.5/206$ \\

\hline
\end{tabular} 
\tablecomments {Model 1: $\tt tbabs\times tbabs\times diskbb$; Model 2: $\tt tbabs\times tbabs\times (diskbb + powerlaw)$; Model 3: $\tt tbabs\times tbabs\times (diskbb + compTT)$; Model 4: $\tt tbabs\times tbabs\times (kerrbb + powerlaw)$.}
\label{chidof}
\end{table*}

\subsection{{\it Suzaku} Observations}

\suzaku{} \citep{Mit07} observed the target multiple times with its on-board instruments and we used the X-ray imaging spectrometer (XIS) data for the analysis. The unfiltered \suzaku{} XIS data were reduced and reprocessed using the specific {\sc headas} tool {\sc aepipeline} in the {\sc heasoft} (version 6.22) software package. The source events were extracted from the circular region of radius 80 arcsec and two source-free circular regions with a radius of 110 arcsec were used to extract background events in XIS 0, 1 and 3. The spectra obtained from XIS 0 and 3, the front-illuminated (FI) CCD spectra, were co-added using the {\sc ftool addascaspec} and the resultant spectra were then grouped with a minimum count of 30--50 per bin using the {\sc grppha} tool. It is noted that the charge leakage area has been increased in XIS 0 after 2015 March 11 and it expanded to the other segments of XIS 0\footnote{\url{http://www.astro.isas.ac.jp/suzaku/doc/suzakumemo/suzaku\_memo\_2015-05.pdf}}. This resulted in the saturation of telemetry and the telemetry saturation periods need to be removed from the observed data conducted after 2015 March 11. Among the {\it Suzaku} observations we used, the S3 observation was performed on 2015 May 18 and we rejected the saturated telemetry data by following the recipe\footnote{\url{http://www.astro.isas.jaxa.jp/suzaku/analysis/xis/xis0\_area\_discriminaion3/Rejecting\_TLMsaturation.pdf}} provided by the instrument team.

\section{Analysis and Results}

Earlier studies suggested a disk-like spectrum for this source \citep{Swa03, Dew13} and we initially used the single-component model, multi-color disk blackbody \citep[{\tt diskbb};][]{Mit84}, to describe the spectra from these observations. The spectral modeling was performed with {\sc xspec} version 12.9.1p \citep{Arn96}. The pn and MOS spectra from \xmm{} were fitted simultaneously in the 0.3--10 keV band, while we used 0.6--10 keV energy range for the \suzaku{} spectra. In the \suzaku{} XIS spectra, we excluded the energy range 1.7--2.0 keV due to the calibration uncertainties. We used two multiplicative absorption components \citep[{\tt tbabs} in {\sc xspec};][]{Wil00} to describe the Galactic absorption towards the direction of the source and intrinsic absorption local to the source. Thus, the first absorption component was fixed at $N_{\rm H,Gal}=5.57\times10^{20}~\rm cm^{-2}$ \citep{Kal05} and second component was considered as a free parameter. The uncertainties on the parameters were quoted at a 90\% confidence level. 

The MCD model provides a satisfactory fit to most of the spectra except X1, X9, S1, and S3. In these observations, the fit provided a reduced $\chi^2$ ($\chi^{2}_{\rm r}=\chi^2/$ degrees of freedom (d.o.f)) of $> 1.6$. In particular, the single component model failed to describe the S3 spectrum, where the $\chi^{2}_{\rm r}$ is 2.9 with 208 d.o.f (see Table \ref{chidof}). In the case of satisfactory fits, the best-fit values of inner disk temperature ($\rm T_{in}$) are in the range of 1.44--1.67 keV. We added a power law (PL) component to the MCD model and this combined model improved the fit for all the spectra, where the improvement in $\chi^2$ varies from $404$ to $\sim 13.6$ for the loss of two d.o.f. The PL index yielded from the fit is $\sim$ 1.65--4.09 and inner disk temperature is roughly consistent with the values obtained from the single-component model. However, in the X9 and S3 observations, we obtained a low temperature of $\sim 0.3$ and 0.8 keV, respectively, compared to the other observations. 

\begin{table*}
\tablecolumns{7}
\setlength{\tabcolsep}{18.0pt}
\tablewidth{320pt}
	\caption{Best-fit parameters for the disk plus Comptonization model}
 	\begin{tabular}{@{}ccccccc@{}}
	\hline
	\hline
\colhead{ObSID} & \colhead{$N_{\rm H}$} & \colhead{$kT_{\rm in}$} & \colhead{$kT_{\rm e}$} & \colhead{$\uptau$} & \colhead{log $L_{\rm X}$} & \colhead{$\chi^2/\rm d.o.f$} \\
\hline

X1 & $0.23^{+0.06}_{-0.05}$ & $0.19^{+0.04}_{-0.03}$ & $1.19^{+0.06}_{-0.06}$ & $11.27^{+1.10}_{-0.91}$ & $39.76^{+0.09}_{-0.06}$ & $465.0/402$ \\
X3 & $0.24^{+0.22}_{-0.16}$ & $0.16^{+0.82}_{-0.16}$ & $1.64^{+0.79}_{-0.50}$ & $>7.02$ & $39.87^{+0.35}_{-0.18}$ & $185.4/193$ \\ 
X4 & $0.19^{+0.29}_{-0.04}$ & $0.13^{+1.02}_{-0.13}$ & $1.59^{+0.91}_{-0.28}$ & $9.65^{+2.67}_{-3.22}$ & $39.74^{+0.44}_{-0.07}$ & $222.9/214$ \\ 
X5 & $0.20^{+0.06}_{-0.02}$ & $0.10^{+0.04}_{-0.04}$ & $1.86^{+0.33}_{-0.19}$ & $8.99^{+0.92}_{-1.05}$ & $39.91^{+0.06}_{-0.02}$ & $277.0/301$ \\

S1 & $0.22(f)$ & $0.14^{+0.01}_{-0.01}$ & $1.29^{+0.05}_{-0.05}$ & $10.00^{+0.52}_{-0.48}$ & $39.96^{+0.03}_{-0.02}$ & $313.5/271$ \\
S2 & $0.22(f)$ & $0.13^{+0.05}_{-0.04}$ & $1.49^{+0.14}_{-0.12}$ & $9.94^{+0.99}_{-0.83}$ & $39.94^{+0.10}_{-0.04}$ & $212.2/230$ \\

X8 & $0.17^{+0.04}_{-0.05}$ & $0.54^{+0.22}_{-0.38}$ & $1.09^{+0.35}_{-0.08}$ & $>10.35$ & $39.79^{+0.10}_{-0.03}$ & $169.0/192$ \\ 
X9 & $0.16^{+0.09}_{-0.05}$ & $0.26^{+0.10}_{-0.08}$ & $>1.82$ & $3.94^{+6.03}_{-1.50}$ & $39.45^{+0.13}_{-0.06}$ & $168.8/193$ \\  
A1 & $0.37^{+0.18}_{-0.18}$ & $0.14^{+0.05}_{-0.05}$ & $1.70^{+1.03}_{-0.32}$ & $8.46^{+2.14}_{-2.59}$ & $40.04^{+0.36}_{-0.10}$ & $178.8/224$ \\   
A2 & $0.20^{+0.26}_{-0.06}$ & $<0.19$ & $1.48^{+0.54}_{-0.22}$ & $9.87^{+1.62}_{-2.51}$ & $39.78^{+0.35}_{-0.06}$ & $224.2/221$ \\ 
S3 & $0.17^{+0.14}_{-0.08}$ & $0.11^{+0.03}_{-0.09}$ & $>2.74$ & $2.42^{+3.29}_{-0.21}$ & $39.89^{+0.23}_{-0.15}$ & $191.1/205$ \\

\hline
\end{tabular} 
\tablecomments {(1) Observation ID; (2) neutral hydrogen column density in units of $10^{22}~\rm cm^{-2}$; (3) inner disk temperature in keV; (4) electron temperature in keV; (5) optical depth (6) logarithmic unabsorbed 0.3--10 keV X-ray luminosity in $\ergsec$, calculated by assuming the distance of 3.63\,Mpc~\citep{Fre94}; (7) $\chi^2$ statistics and degrees of freedom.}
\label{comptt}
\end{table*}

Many ULXs show peculiar features in the high quality {\it XMM-Newton} spectra: a rollover at high energy, usually at $\sim 3-5$ keV, often coupled with a soft excess \citep{Sto06, Gla09}. Such spectral features can be described by the disk plus Comptonized corona model. To study the spectral evolution of M81 X--6, we used the disk plus Comptonized corona model, namely {\tt diskbb} + {\tt comptt} \citep{Tit94} model. In the model fit, we set the seed photon temperature ($\rm T_{0}$) equal to the inner disk temperature ($\rm T_{in}$). This model improved the spectral fit for the X1 spectrum compared to the simple empirical model, MCD plus PL. The difference in the $\chi^2$ value for X1 is $\sim 34.7$ for the loss of one additional d.o.f. For the rest of the data, this model fit is comparable to the MCD plus PL model. In some of the {\it Suzaku} observations (S1 and S2), we observed the degeneracy between the neutral hydrogen column density and disk blackbody model parameters. In order to mitigate the degeneracy, we fix the absorption column density at an average value ($N_{\rm H}=2.20\times10^{21}~\rm cm^{-2}$) obtained from the {\it XMM-Newton} data. The derived values of the inner disk temperature are in the range of $\sim 0.1-0.5$ keV. The plasma temperature obtained from the spectral fit is not well constrained in some observations and it is in the range of $1.1-1.9$ keV, while the values of optical depth are $\sim 2.4-11.3$. The best-fit spectral model parameters are given in Table \ref{comptt}. 

To further investigate the spectral evolution of M81 X--6, we tested a more sophisticated accretion disk model, instead of MCD model, along with PL component. MCD is a simple model and it neglects all the relativistic effects. Since the source is disk-dominated, it is important to investigate the properties with more sophisticated disk model. Thus, we replaced the simple MCD with {\tt kerrbb} \citep{Li05} model in the two-component model (MCD plus PL), which provides the black hole mass, accretion rate and spin of the system in a self-consistent manner. {\tt kerrbb} is the multi-temperature blackbody model for a thin, steady-state accretion disk around a Kerr black hole, which incorporates the general relativistic effects of spinning black hole. This model also has taken into account the effect of self-irradiation of the disk, limb darkening, and the torque at the inner boundary of the disk. We set the torque at the inner boundary of the disk as zero and switch on the self-irradiation and limb darkening flag of the model. The disk inclination angle is fixed at $60^{\circ}$, the spectral hardening factor is assumed to be 1.7 \citep{Shi95} and fixed at this value.

\begin{table*}
\tablecolumns{7}
\setlength{\tabcolsep}{17.0pt}
\tablewidth{320pt}
	\caption{Best-fit parameters for {\tt kerrbb} plus PL model}
 	\begin{tabular}{@{}cccccccc@{}}
	\hline
	\hline
\colhead{ObSID} & \colhead{Case} & \colhead{$N_{\rm H}$} & \colhead{$\dot{M}$} & \colhead{$\Gamma$} & \colhead{$\chi^2/\rm d.o.f$} & \colhead{ER} & Group \\
\hline
$\rm X1^{*}$ & I   & $0.47^{+0.04}_{-0.05}$ & $78.38^{+1.79}_{-1.76}$ & $4.31^{+0.24}_{-0.27}$ & $511.1/404$ & 5.60 & M \\
   & II  & $0.18^{+0.02}_{-0.01}$ & $23.71^{+1.92}_{-1.97}$   & $2.22^{+0.19}_{-0.12}$ & $557.4/404$ & 0.53 & \\
   & III & $0.17^{+0.01}_{-0.02}$ & $13.94^{+1.13}_{-1.14}$   & $2.15^{+0.16}_{-0.11}$ & $547.9/404$ & 0.16 & \\

X3 & I   & $0.13^{+0.09}_{-0.05}$ & $37.55^{+30.58}_{-37.55}$ & $1.39^{+0.20}_{-1.39}$ & $192.5/195$ & 2.68 & M \\
   & II  & $0.12^{+0.07}_{-0.03}$ & $15.92^{+7.73}_{-15.92}$  & $1.30^{+0.30}_{-1.30}$ & $192.8/195$ & 0.36 & \\
   & III & $0.11^{+0.07}_{-0.02}$ & $10.32^{+5.60}_{-7.72}$ & $1.22^{+0.34}_{-1.22}$ & $192.4/195$ & 0.12 & \\

X4 & I   & $0.12^{+0.05}_{-0.05}$ & $46.84^{+25.61}_{-21.38}$ & $1.48 ^{+0.17}_{-0.37}$ & $223.6/216$ & 3.35 & M \\
   & II  & $0.11^{+0.06}_{-0.03}$ & $20.77^{+7.42}_{-9.85}$   & $1.31^{+0.28}_{-1.31}$ & $223.8/216$ & 0.46 & \\
   & III & $0.10^{+0.06}_{-0.02}$ & $12.88^{+3.17}_{-6.21}$ & $1.21^{+0.35}_{-1.21}$ & $223.2/216$ & 0.15 & \\

X5 & I   & $0.16^{+0.02}_{-0.02}$ & $44.95^{+17.99}_{-16.43}$ & $1.46^{+0.07}_{-0.09}$ & $277.3/303$ & 3.21 & M \\
   & II  & $0.14^{+0.03}_{-0.03}$ & $20.14^{+8.19}_{-7.26}$   & $1.38^{+0.11}_{-0.16}$ & $276.7/303$ & 0.45 & \\
   & III & $0.14^{+0.03}_{-0.03}$ & $12.56^{+5.60}_{-4.62}$ & $1.34^{+0.13}_{-0.24}$ & $276.1/303$ & 0.14 & \\

S1 & I   & $0.13^{+0.08}_{-0.05}$ & $89.97^{+4.79}_{-6.84}$   & $2.96^{+0.54}_{-0.43}$ & $322.7/272$ & 6.43 & H \\
   & II  & $0.18^{+0.13}_{-0.15}$ & $37.29^{+1.24}_{-3.34}$   & $3.61^{+0.66}_{-1.20}$ & $321.4/272$ & 0.83 & \\
   & III & $0.05^{+0.24}_{-0.04}$ & $21.04^{+1.54}_{-1.43}$ & $2.65^{+1.54}_{-0.44}$ & $321.7/272$ & 0.24 & \\

S2 & I   & $0.04^{+0.06}_{-0.02}$ & $101.54^{+18.94}_{-20.79}$ & $1.45^{+0.30}_{-1.45}$ & $211.4/231$ & 7.26 & H \\
   & II  & $0.05^{+0.02}_{-0.02}$ & $40.90^{+3.03}_{-2.49}$   & $1.00^{+0.48}_{-1.00}$ & $212.5/231$ & 0.91 & \\
   & III & $0.06^{+0.23}_{-0.03}$ & $24.10^{+2.63}_{-5.43}$ & $0.78^{+0.85}_{-1.85}$ & $212.9/231$ & 0.28 & \\

X8 & I   & $0.16^{+0.04}_{-0.04}$ & $68.32^{+29.64}_{-25.73}$ & $1.70^{+0.43}_{-0.18}$ & $183.2/194$ & 4.88 & M \\
   & II  & $0.14^{+0.05}_{-0.04}$ & $27.77^{+11.61}_{-10.73}$ & $1.53^{+0.20}_{-1.53}$ & $184.2/194$ & 0.62 & \\
   & III & $0.13^{+0.05}_{-0.03}$ & $17.44^{+6.01}_{-6.79}$ & $1.43^{+0.25}_{-1.43}$ & $182.9/194$ & 0.20 & \\

X9 & I   & $0.19^{+0.02}_{-0.03}$ & $ 0.39^{+2.08}_{-0.39}$   & $2.22^{+0.10}_{-0.11}$ & $188.0/195$ & 0.03 & L \\
   & II  & $0.19^{+0.02}_{-0.03}$ & $ 0.16^{+0.88}_{-0.16}$   & $2.22^{+0.09}_{-0.11}$ & $188.0/195$ & 0.004 & \\
   & III & $0.19^{+0.02}_{-0.03}$ & $ 0.12^{+0.48}_{-0.12}$ & $2.21^{+0.10}_{-0.10}$ & $188.0/195$ & 0.001 & \\

A1 & I   & $0.17^{+0.03}_{-0.03}$ & $42.39^{+22.22}_{-19.73}$ & $1.67^{+0.11}_{-0.11}$ & $184.3/226$ & 3.03 & M \\
   & II  & $0.16^{+0.04}_{-0.04}$ & $17.49^{+9.29}_{-8.28}$   & $1.60^{+0.12}_{-0.18}$ & $184.9/226$ & 0.39 & \\
   & III & $0.16^{+0.04}_{-0.04}$ & $10.51^{+5.82}_{-4.98}$ & $1.58^{+0.13}_{-0.23}$ & $184.4/226$ & 0.12 & \\

A2 & I   & $0.16^{+0.05}_{-0.05}$ & $50.76^{+26.40}_{-22.80}$ & $1.66^{+0.20}_{-0.20}$ & $223.2/223$ & 3.63 & M \\
   & II  & $0.14^{+0.05}_{-0.05}$ & $22.18^{+10.57}_{-9.98}$  & $1.53^{+0.20}_{-1.53}$ & $223.1/223$ & 0.50 & \\
   & III & $0.13^{+0.06}_{-0.04}$ & $13.33^{+6.21}_{-6.15}$ & $1.48^{+0.23}_{-1.48}$ & $223.0/223$ & 0.15 & \\

S3 & I   & $0.07^{+0.04}_{-0.04}$ & $11.55^{+8.30}_{-5.77}$   & $1.96^{+0.09}_{-0.08}$ & $190.2/207$ & 0.83 & L \\
   & II  & $0.07^{+0.05}_{-0.04}$ & $ 4.88^{+3.56}_{-2.44}$   & $1.94^{+0.10}_{-0.09}$ & $190.2/207$ & 0.11 & \\
   & III & $0.07^{+0.05}_{-0.04}$ & $ 2.76^{+2.04}_{-1.38}$ & $1.94^{+0.10}_{-0.09}$ & $190.2/207$ & 0.03 & \\

\hline
\end{tabular} 
\tablecomments {(1) Observation ID; (2) different spin and mass cases, case I: $a^{*}=0$ \& $\rm M=10\,\rm M_\odot$, case II: $a^{*}=0.9$ \& $\rm M=32\,\rm M_\odot$, case III: $a^{*}=0.999$ \& $\rm M=62\,\rm M_\odot$; (3) neutral hydrogen column density in units of $10^{22}~\rm cm^{-2}$; (4) mass accretion rate in $10^{18}~\rm g~s^{-1}$; (5) photon index; (6) $\chi^2$ statistics and degrees of freedom (7) Eddington Ratio; (8) accretion rate group, where L, M and H are low, medium and high accretion rate, respectively. $^{*}$ indicates the spectral fits obtained from the {\tt kerrbb} plus PL model is significantly worse compared to the disk plus Comptonization model (see the text for more details).}
\label{all}
\end{table*}

Initially, we fitted the observed spectra by considering the spin, mass and accretion rate as free parameters. The fit is formally acceptable in the majority of the cases, except for observation X1, that is not satisfactorily fitted with the {\tt kerrbb} plus PL model ($\rm \chi^2/d.o.f=508.6/402$). Therefore, either the latter model does not apply to this source or it does not provide an adequate representation of all its spectral variability. While we are aware of this, in the following we assume that the {\tt kerrbb} plus PL model can correctly represent the physical state of at least part of the X-ray observations of M81 X--6 and use it in an attempt to constrain its physical parameters (mass, spin and accretion rate). Leaving all the parameters free to vary gives completely unconstrained values of the BH specific angular momentum. So, we first fixed it at three representative values (0, 0.9 and 0.999) and repeated the spectral fits considering the mass and accretion rate as free parameters. Since the mass of the BH should not change, we then fixed it at the average value derived from all the fits with a given BH specific angular momentum. The inferred values are: $\sim 10\,\rm M_\odot$ ($a^{*}=0$), $32\,\rm M_\odot$ ($a^{*}=0.9$), $62\,\rm M_\odot$ ($a^{*}=0.999$), respectively. At this point, we fixed the BH mass and spin, and repeated the spectral fitting. The best-fit spectral parameters for three spin values are given in Table \ref{all}. The best fit obtained for the three cases (fixed values of mass and spin) are comparable. However, as shown in Table \ref{all}, the Eddington ratios for the majority of the fits with a non-rotating black hole (case I) are larger than unity. As the adopted model is no longer consistent in this regime \citep[see][]{Hui08}, the physical parameters inferred from them are not reliable, and we will not consider them further. The PL indices are in the range, $\Gamma \sim 1.0 - 2.2$, with the exception of observation S1 ($\Gamma \sim 2.6 - 3.6$). We also note that the soft end of these spectra is dominated by the steep PL continuum, as in NGC 5204 X--1, M33 X--8, NGC 55 ULX \citep{Fos04, Sto04, Rob05, Gla09}, indicating that also this observation (as X1) is likely not to be consistent with this model.

\begin{figure}
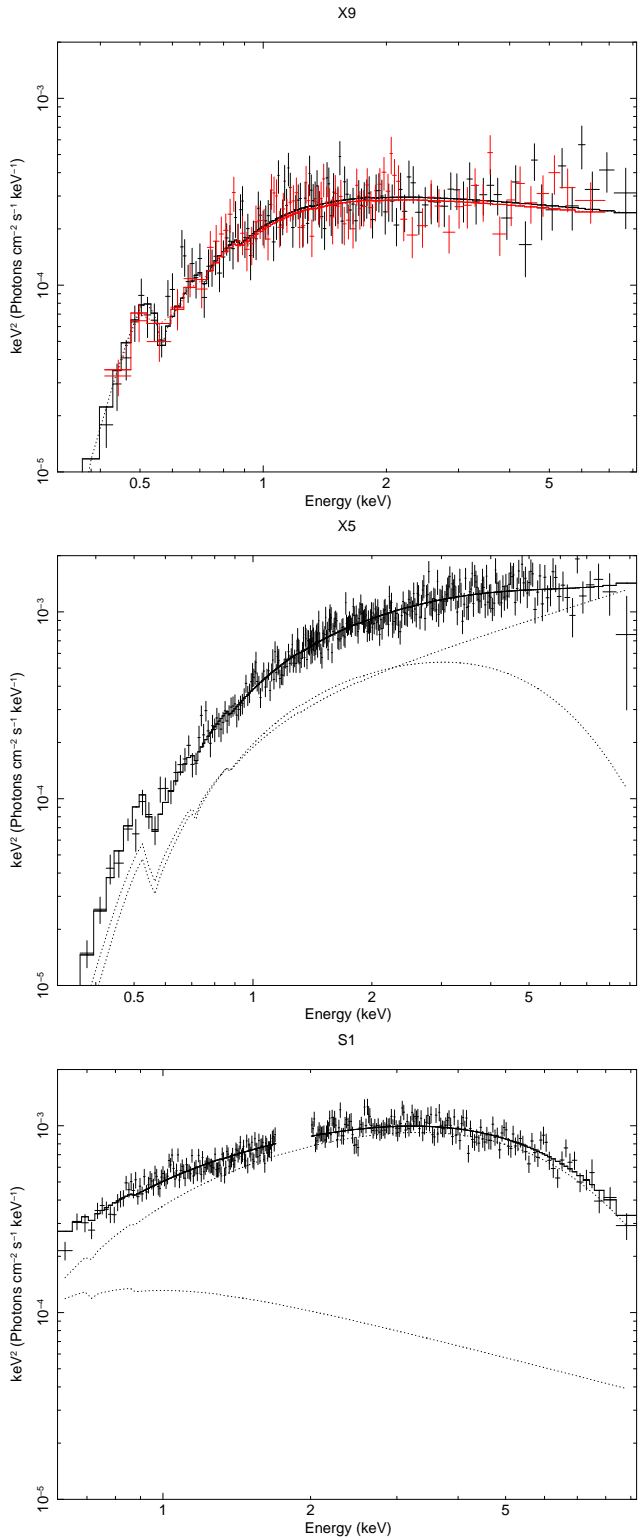

\begin{center}

 \includegraphics[width=6.8cm,angle=-90]{f1a.ps}
 \includegraphics[width=6.8cm,angle=-90]{f1b.ps}
 \includegraphics[width=6.8cm,angle=-90]{f1c.ps}

\caption{The three spectral shapes of M81 X--6 observed in the \xmm{} and \suzaku{} observations. Spectra showed here are from the X9 (top), X5 (middle) and S1 (bottom) observations fitted with {\tt kerrbb} plus PL model. In \xmm{} observations, the black, and red data points represent the pn and MOS data, respectively.}
\label{shape}
\end{center}
\end{figure}

\begin{figure}
\begin{center}

\includegraphics[width=8.9cm,angle=0]{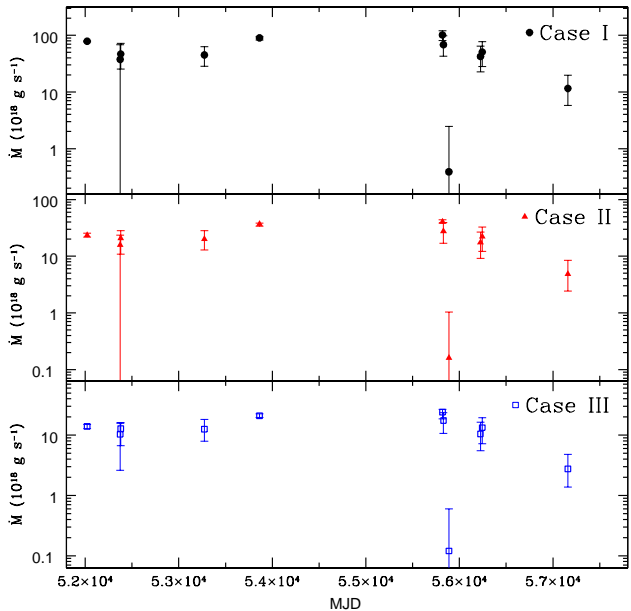}
\caption{Variation of mass accretion rate as a function of time for three cases.}
\label{rate}
\end{center}
\end{figure}

Based on the mass accretion rate derived using the {\tt kerrbb} plus PL model, we classify the observations into three categories: low, medium and high accretion rate groups. The observations X9 and S3 belong to the low accretion state, while the S1 and S2 data in the high accretion rate. The rest of the data can be classified into medium accretion rate level. In these three accretion rates, the source exhibits different spectral shapes as shown in Figure \ref{shape}. The derived accretion rate from the model is varied by more than an order of magnitude in these observations as shown in Figure \ref{rate}. The accretion rate seems to be constant within the uncertainties during the first four observations. It shows an increasing trend in the subsequent observations and then decreases to reach lowest accretion rate in 2011 November \xmm{} observation. But later observations reveal that the accretion rate gradually increases once again and then decreases to the low accretion level in the last observation. The observed trends in the mass accretion rate and different spectral shapes indicate the variability of the source. Variation in the PL index is less significant compared to the accretion rate in these observations. It is also noted that the PL indices have similar values in the low accretion rate observations, while in the high accretion rate data the indices are different, i.e., soft and hard PL index in the high luminosity. We also investigated the parameter correlations and unfortunately, we do not find any significant correlations between the spectral parameters.     

\section{Discussion}

We analyzed the \suzaku{} and \xmm{} observations of the ULX M81 X--6 conducted between 2001 and 2015, to study the spectral variability of the source. We initially used the disk plus Comptonized corona model to investigate the spectral evolution of the source. Such disk-corona models were widely used to describe the observed spectra of ULXs \citep{Sto06, Gla09, Pin12, Jit17}. For the majority of ULXs, such model fits provide a cool plasma temperature ($\sim 1-4$ keV) and optically thick corona ($\uptau \sim 5-30$), unlike the $\uptau \sim 1$ coronae seen in BH X-ray binaries, \citep{Sto06, Gla09, Pin14}. In our modeling with {\tt diskbb} plus {\tt comptt} components, the yielded spectral parameters are consistent with the other ULXs. The optical depth obtained for the X9 and S3 observations is small, $\uptau \sim 2-4$, which indicates that the source may belong to the ``thick corona'' state ($\uptau \lesssim 8$) in these observations as explained in \cite{Pin12}. Moreover, in the other observations that we analyzed, the obtained values of optical depth are in the range of $\uptau \sim 8-11$, which indicates a ``very-thick corona'' state for the source. Thus, the different values of optical depth may represent physically different states for the source.

We also used the two-component spectral model, relativistic accretion disk ({\tt kerrbb}) plus PL to explain the observed spectra of the source and systematically studied the spectral properties. Our analysis revealed that the mass accretion rate is varied by more than an order of magnitude during these observations and exhibited different spectral shapes. The observed spectral shapes are related to the accretion rate of the system. At lower accretion rate, the source tends to show PL-dominated spectrum, while as the accretion rate increases (in the medium and high accretion rates) the source exhibits the disk-dominated spectrum. However, observation X1 is not satisfactorily fitted with the {\tt kerrbb} plus PL model, implying that this model does not provide a full representation of its 
spectral variability. 

Although Eddington ratios larger than unity are physically possible for BHs accreting above Eddington, they require an accretion disk with a different physical structure \citep[e.g., ][]{Abr88, Wat00, Beg02}. The results of the spectral fits of the {\tt kerrbb} plus PL model for a non-rotating BH provide Eddington ratios in excess of one and are thus inconsistent with the assumptions of the model. For this reason, we did not consider them further. However, for rapidly rotating BHs ($a^{*}=0.9$ and 0.999), the fits return physically consistent values of the parameters. The average value of the inferred BH mass is in the range of MBHs ($\rm M < 100\,\rm M_\odot$). In these cases, the source can emit at near the Eddington limit.

That the inferred mass depends on the assumed value of the spin has been known for other ULXs \citep{Fen10, Str14, Bri16}. For example, analysis of ULX M82 X--1 spectra using \chandra{} and \xmm{} observations suggest a StMBH system for the source accreting at highly super-Eddington limit ($\sim 160$) for a low spin value $a^{*}=0$, while for the maximally spinning case ($a^{*}=0.9986$) the inferred mass is in the range of $\sim 200-800\,\rm M_\odot$, consistent with IMBHs \citep{Fen10}. The broadband study of M82 X--1 using the \swift{}, \chandra{} and \nustar{} observations estimates the mass of the system as $\sim 26\,\rm M_\odot$ with zero spin, while an IMBH of mass $\sim 125\,\rm M_\odot$ is inferred when assuming maximal spin \citep{Bri16}. Thus, the spin is an important parameter that clearly has an impact on the accretion scenarios (Eddington or Super-Eddington) of ULXs.

Previous studies estimated the mass of M81 X--6 using different model assumptions \citep{Mak00, Swa03, Hui08, Dew13}. For example, assuming the innermost disk radius derived from the disk blackbody model, \citet{Swa03} derived the mass of the compact object as $\sim 18\,\rm M_\odot$, which is consistent with the case of non-spinning StMBHs. Moreover, the mass derived from the two relativistic disk models, {\tt kerrbb} and {\tt bhspec}, using single \xmm{} observation of M81 X--6 spans in the range of $33-85\,\rm M_\odot$ \citep{Hui08}. The mass inferred from our study using high spin values is consistent with the previous studies and clearly ruling out the IMBH accretor for M81 X--6. Thus, the source may harbor MBH and accrete at near Eddington limit. It is interesting to note that the estimated mass of M81 X--6 from all the studies is consistent with the masses of merged BHs discovered through the gravitational wave detection, where the final mass of BH is in the range of $\sim 18-62\,\rm M_\odot$ \citep[][and references therein]{Abb16, Abb16a, Abb17b} in the five gravitational wave detection events. This confirms the presence of MBHs in the Universe and provides a possible explanation for the formation mechanism of such systems. 

It is also interesting to consider the multi-wavelength properties of M81 X--6. In the optical wavelength, a counterpart has been identified for M81 X--6 in the {\it HST} image and the inferred optical properties are consistent with an early-type main-sequence star of spectral class O8 V \citep{Liu02}. Neglecting binary evolution effects the estimated mass of the companion is $23\,\rm M_\odot$. Using the {\it HST} images, \citet{Swa03} confirmed that the optical counterpart has magnitude and colors resembling those of an early-type main-sequence star and classified it into the O9-B1 spectral class. Thus, M81 X--6 most likely belongs to the HMXB category. Since the donor is a main-sequence star, the stellar wind from the system may not be strong as compared to the supergiant's and the compact object accretes the matter via Roche lobe overflow. By assuming the companion star of mass $23\,\rm M_\odot$ and compact object of $18\,\rm M_\odot$, \citet{Liu02} estimated the separation between the primary and secondary stars, which is $\sim 1.5 \times 10^{12}~\rm cm$. However, for the mass of the compact object obtained from this study, $32-62\,\rm M_\odot$, the expected separation ranges from $1.7-2.0 \times 10^{12}~\rm cm$. Such a separation provide the space to form a large accretion disk around the black hole and the large disk can act as a huge reservoir of accreted mass \citep{Tru06, Dee09}. Thus, the observed long-term activity of M81 X--6 can be associated with the large orbital separation and consequently the accreted mass in the disk.

The orbital separation estimation also predicts the orbital period of the binary system, which is $\sim 1.8$\,d \citep{Liu02}. However, such an orbital period has not been identified for the source, instead, two variable and statistically less significant ($< 99\%$) peaks near 110\,d and 370\,d have been detected in the LSP, which are likely the super-orbital period of the system \citep{Lin15}. One of the possible reason for the long-term X-ray quasi-periodic modulations is the mass transfer rate related events such as X-ray state transition and the disk instability \citep[see][and references therein]{Kot12}. Alternatively, this can also be explained by the tidal interaction-induced disk precession \citep{Whi91} and the radiation-driven warping of accretion disks \citep{Ogi01}.

We note that some of the ULXs have shown pulsations which reveal that they are NS system \citep{Bac14, Isr17a, Isr17b}. For M81 X--6, no pulsation has been reported so far. Hence, it is still not certain whether it is an accreting NS system. Indeed, ULXs may be an heterogeneous system consisting of both NS or BH. Our results should be interpreted on the assumption that M81 X--6 is an accreting BH system. 
 
\section{Summary}

In this paper we have presented the analysis of \suzaku{} and \xmm{} observations of the ULX M81 X--6 during the period 2001 and 2015. We initially fitted the observed spectra with disk plus Comptonized corona model and the inferred spectral parameters ($\rm kT_{e} \sim 1-2$ keV and $\uptau \sim 2-11$) are consistent with other ULXs. Such a model fit revealed the spectral variability of the source with a change in the optical depth of the Comptonizing component. We further investigated the spectral evolution of M81 X--6 using the relativistic accretion disk emission plus a power law component. While the disk plus thermal Comptonization model can adequately reproduce all the observations, the {\tt kerrbb} plus PL model can reproduce a large part but not all of the spectral variability of the source. The source exhibits three different spectral shapes during these observations and the observed spectral shapes are associated with the accretion rate of the system. The systematic study by fixing the spin and mass of the system for all observations rule out IMBH ($M_{\rm BH} > 10^{2}-10^{4}\,\rm M_\odot$) compact object for M81 X--6, instead our study suggests a rapidly spinning massive BH ($20-100\,\rm M_\odot$). The inferred accretion rates, in that case, are consistent with the near Eddington limit. Since there is no pulsation detected from this source so far, we interpreted our results assuming an accreting BH X-ray binary system. However, the future dedicated observations are required to confirm the real nature (BH or NS system) of the source.

\section*{Acknowledgements}

We thank the anonymous referee for the constructive comments and suggestions that significantly improved this manuscript. This research has made use of data obtained from the High Energy Astrophysics Science Archive Research Center (HEASARC), provided by NASA's Goddard Space Flight Center. \\

{\it Software}: HEAsoft \citep[v 6.22;][]{Arn96}, SAS \citep[v 14.0][]{Gab04}, XSPEC (v 12.9.1p).


\end{document}